\documentclass[aps,prl,twocolumn,superscriptaddress]{revtex4}
\usepackage{amsmath}
\usepackage{amssymb}
\usepackage[dvips]{graphicx}
\usepackage{wrapfig}
\usepackage{subfigure}
\usepackage{psfrag}

\begin{document}
\DeclareGraphicsExtensions{.eps}
\title{Gapless spin liquids on the three dimensional hyper-kagome 
lattice of Na$_4$Ir$_3$O$_8$}
\author{Michael J. Lawler}
\author{Arun Paramekanti}
\author{Yong Baek Kim}
\affiliation{Department of Physics, University of Toronto, Toronto, Ontario M5S 1A7, Canada} 
\author{Leon Balents}
\affiliation{Kavli Institute for Theoretical Physics, University of California, 
Santa Barbara, CA 93106}

\date{\today}

\begin{abstract}
Recent experiments indicate that Na$_4$Ir$_3$O$_8$, a 
material in which $s\!\!=\!\!1/2$ Ir local moments form a three dimensional 
network of corner-sharing triangles, may have a quantum spin liquid ground
state with gapless spin excitations. Using a combination of exact 
diagonalization, symmetry analysis of fermionic mean field ground states 
and Gutzwiller projected variational wavefunction studies, we propose
a quantum spin liquid with spinon Fermi surfaces as a favorable 
candidate for the ground state of the Heisenberg 
model on the hyper-kagome lattice of Na$_4$Ir$_3$O$_8$. 
We present a 
renormalized mean field theory of the specific heat of this 
spin liquid and also discuss possible low temperature instabilities
of the spinon Fermi surfaces.
\end{abstract}
\maketitle

\begin{figure}[t]
\includegraphics[width=0.4\textwidth,height=0.3\textwidth]{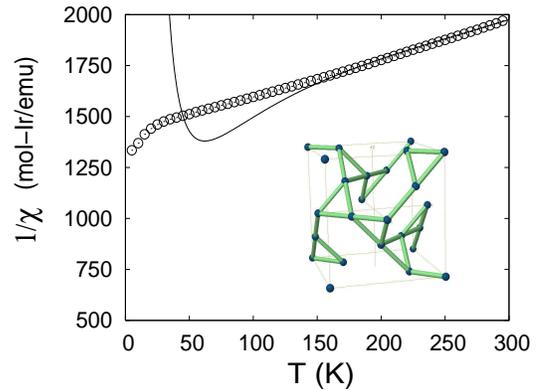}
\caption{Exact diagonalization (ED) results [solid line] for the inverse 
uniform magnetic susceptibility $1/\chi$ 
compared with experiments [open 
circles]. The ED was done on a single unit cell of the
hyper-kagome lattice (inset) with $J=304K$ chosen to reproduce the
high temperature experimental $1/\chi$.}
\label{Fig:ED}
\end{figure}

\noindent {\bf Introduction:}
Na$_4$Ir$_3$O$_8$ is a recently discovered three dimensional (3D) frustrated
quantum magnet \cite{takagi}. The Ir atoms in this insulating compound have 
$s\!\!=\!\!1/2$
local moments and form a 3D network of corner sharing triangles called a 
`hyper-kagome' lattice \cite{takagi}, 
a cubic lattice whose unit cell is shown in 
Fig.~\ref{Fig:ED}. High temperature magnetic susceptibility ($\chi$)
measurements in this material suggest that the Ir moments have strong 
antiferromagnetic correlations with a Curie-Weiss temperature $\Theta_W
\sim -650 K$. The observation of a large $\chi$ and entropy
at low temperature indicates that gapless spinful excitations survive for
$T \ll \Theta_W$.
At the same time, $\chi$ and specific heat 
measurements reveal no sign of any magnetic order or any other symmetry
breaking down to $T \sim 0.5K$, nearly three orders of magnitude 
lower than $\Theta_W$, suggesting that
Na$_4$Ir$_3$O$_8$ may be the first
example of a 3D quantum spin liquid which does not order
down to $T\!=\!0$. It joins a small but growing list of recently
discovered frustrated $s\!\!=\!\!1/2$ quantum magnets \cite{spinliquid} which appear to have
quantum disordered ground states possibly supporting fractionalized
excitations.

These experiments motivated a study of the classical Heisenberg
antiferromagnet on the hyper-kagome lattice \cite{hopkinson}. This model
was found to order into a coplanar `classical nematic' state at low
temperatures, $T \lesssim J/1000$, where $J$ is the nearest neighbor
antiferromagnetic exchange coupling. However, quantum effects are
clearly significant at such low temperatures. A subsequent study of the
quantum Heisenberg model, using an Sp(N) mean field theory, uncovered a
candidate quantum spin liquid ground state with $Z_2$ topological order
as well as an interesting magnetically ordered state which is proximate
to this spin liquid \cite{lawler}. However, this `bosonic' spin liquid
has a nonzero spin gap which is at odds with recent observations, that
gapless spin excitations survive down to $T \sim 0.5K$ \cite{takagi2},
unless the spin gap is anomalously small.  Another difficulty of this
proposal is that there should be a finite temperature transition from
the $Z_2$ spin liquid to the higher temperature paramagnetic phase while
there is no clear signature of a phase transition in thermodynamic
measurements \cite{takagi}.

Here we pursue a completely different line of attack and attempt to
build a `fermionic' spin liquid theory of the hyper-kagome Heisenberg
model. This formulation has the virtue that gapless spin liquids emerge
as stable phases at mean field level and beyond without any need for
fine tuning \cite{xgwen,ran}.  The main results of our paper are as
follows. (i) We find that of a number of candidate spin liquid ground
states we have explored, a particularly simple fermionic spin liquid
state, one which supports Fermi surfaces of spinons, emerges as a
promising candidate for the ground state of the nearest neighbor
Heisenberg model on the hyper-kagome lattice. This result is obtained by
a combination of exact diagonalization, a projective symmetry group
(PSG) analysis \cite{xgwen} of mean field ground states, and Gutzwiller
projected variational wavefunction calculations. (ii) We then show,
using a Gutzwiller renormalized mean field theory \cite{rmft}, that the
specific heat of this spin liquid state is quite similar to the
experimentally observed specific heat of Na$_4$Ir$_3$O$_8$ for $T
\gtrsim 5K$.  This spinon Fermi surface state therefore seems to be a
good starting point to understand the physics of this material over a
wide range of temperatures in the same way that Fermi liquid theory is a
good starting point to understand conventional metals. However, as in
conventional metals, the Fermi surfaces could be unstable, at very low
temperature, due to small additional interactions.  (iii) 
We considered a symmetry analysis of possible low
temperature instabilities of the spinon Fermi surface state. 
Some of the states resulting from this analysis support line nodes 
for spinon excitations.
We discuss implications of such line node states for the specific heat 
data.

%\bigskip

\noindent {\bf Model and exact diagonalization:}
We begin with an exact diagonalization (ED) study of the nearest neighbor
$s=1/2$ Heisenberg model 
\begin{equation}
H = \frac{1}{2}\sum_{ij} J_{ij} \vec S_i \cdot\vec S_j,
\end{equation}
on a single 12-site unit cell of the
the hyper-kagome lattice formed by the Ir sites in Na$_4$Ir$_3$O$_8$
(see inset of Fig.~\ref{Fig:ED}).
Here $J_{ij}$ is the exchange coupling on the bond $ij$, and we keep
only the nearest neighbor antiferromagnetic exchange interaction, $J > 0$.
Fig.~\ref{Fig:ED} displays the ED result for $\chi^{-1}(T)$ with a choice
of $J=304K$
which, as shown, reproduces the experimental data
at high temperatures. We find that for $T=200-300 K$, the $\chi^{-1}(T)$ 
from ED can be fit by a ``Curie-Weiss'' law with $\Theta_{W} \approx -730K$.
This is only an apparent Curie-Weiss behavior in the sense that
it is only valid for a limited range of temperature around $J$
and not for $T \gg J$ where one recovers the usual Curie-Weiss law
with $\Theta_W^{T \gg J} = - z J /4$, with the coordination number $z=4$ 
on the hyper-kagome lattice. The upturn in
$\chi^{-1}(T)$ in the ED for $T \lesssim 50K$ arises from a 
nonzero spin gap on a single unit cell. 

Experiments observe a broad peak in $C/T$
at $T_p \approx 25 K$ with a peak height $(C/T)_{\rm max} \approx 
55$mJ/K$^2$/mol-Ir \cite{takagi}. While finite size effects are clearly important in
the ED calculations at low temperatures, it is nevertheless encouraging 
that the ED result for $C/T$ of the Heisenberg model with $J=304K$ 
shows a broad peak at $T_p \approx 20K$ with a
peak height of $(C/T)_{\rm max} \approx 70$mJ/$K^2$/mol-Ir.

Since the $s=1/2$ Heisenberg model appears to capture some aspects of
the experimental data on Na$_4$Ir$_3$O$_8$, we now turn to an analysis of 
possible fermionic spin liquid candidates for the ground state of this 
model in order to see if we can understand the emergence of a gapless
spin liquid ground state in Na$_4$Ir$_3$O$_8$.

%\bigskip

\noindent{\bf Hyper-kagome spin liquid states:}
We begin by representing spin operators
in terms of fermionic spinors with the constraint of one fermion per
site. The Hamiltonian $H$ can then be decoupled in both the hopping
(Hartree-Fock) and pairing (Bogoliubov) channels within a mean field
approximation. This decoupling leads to the mean field Hamiltonian
\begin{multline}
  H_{MF} = \frac{3}{8}\sum_{\langle ij\rangle} J_{ij} \bigg[
  \frac{1}{2}\text{Tr} \big({\bf U}_{ij}^\dagger\cdot {\bf U}_{ij} \big)
  - \Psi_i^\dagger\cdot {\bf U}_{ij}\cdot\Psi_j + h.c.\bigg] \\+ \sum_i
  \vec a_i\cdot \Psi^\dagger_i\cdot\vec\tau\cdot\Psi_i
\end{multline}
written in a manifestly SU(2) invariant form where $\Psi_i^T = (f_{i\uparrow}, f^\dagger_{i\downarrow})^T$ is a Nambu spinor,
\begin{equation}
  {\bf U}_{ij} = \begin{pmatrix} \chi_{ij} & \Delta_{ij} \\ \Delta^*_{ij} & -\chi^*_{ij} \end{pmatrix}
\end{equation}
are the mean field hopping and pairing amplitudes and $\vec a_i$ is a
lagrange multiplier which \emph{on average} enforces, in an SU(2)
invariant manner, the single occupancy constraint $\langle
f^\dagger_{i\sigma}f_{i\sigma}\rangle = 1$.

We seek the ground state of $H$. Guided by experiment, assume the ground
state is a spin liquid, that it does not break any lattice symmetries,
global spin-rotation symmetry or time-reversal symmetry.  Remarkably,
even aside from the known SU(2) gauge redundancy, imposing all these
symmetries on the mean field ground state does not single out a unique
choice for $U_{ij}$ and $\vec a_i$. The identification of non-redundant
spin liquid phases, therefore, requires careful examinations of the
combined space group and gauge transformations. These fundamental
symmetry relations, proposed by Wen \cite{xgwen} and dubbed ``projective
symmetry group (PSG)'' can be used to classify all possible symmetric
spin liquid phases.
Under a PSG transformation, spinors transform as $\Psi_i \to {\bf
  G}^X_i\cdot\Psi_{X(i)}$ where ${\bf G}^X_i$ is an SU(2) gauge
transformation associated with the space group transformation
$X$. Transforming a mean field ansatz by:
\begin{equation}
  U_{ij} \to {\bf G}^X_i\cdot U_{X(i)X(j)}{\bf G}^X_j,\quad 
  \vec a_i\cdot\tau \to {\bf G}^X_i(\vec a_{X(i)}\cdot\tau){\bf G}^X_i
\end{equation}
then leaves $H_{MF}$ invariant. Naturally, these PSGs are subgroups of
the combined space and gauge groups, which leads to strong constraints on
the possible choices of ${\bf G}^X_i$.

We have constructed a systematic classification of spin liquid ground
states by constructing all PSGs with non-trivial ${\bf G}^X_i$
associated with the \emph{point group} of the hyper-kagome lattice. This
group turns out to be equivalent to the octahedral group $O$ and
consists of two-fold rotations about each site, three-fold rotations for
each triangle and four-fold screw rotations for each thread (see
Ref. \cite{lawler} for a discussion of threads). Details of our
calculation will be presented in a future publication
\cite{lawler2}. Here, for simplicity, we focus on the family of states
\begin{equation}\label{eq:ansatz}
      U_{ij} = \chi_{ij} \tau_3 + \Delta_{ij} \tau_1, \quad \vec a_i = -\mu \hat z
\end{equation}
where $\chi_{ij}$ is real and positive, $\Delta_{ij}$ is real but
alternates sign as discussed below and only the bonds $ij$ that have a
finite exchange $J_{ij}$ have finite $U_{ij}$. This family of states
covers most of the states resulting from our PSG analysis.

From a symmetry perspective, $\chi_{ij}$ and $\Delta_{ij}$ are chosen to
be invariant under translations and three-fold rotations through each
triangle. However, two-fold rotations about each site and the four-fold
screw rotations both need to be followed by the gauge transformation
${\bf G}^X_i = i\tau_3$, where $X$ is either of these
transformations. This second requirement fixes the sign of the pairing
fields $\Delta_{ij}$. In addition to these spatial symmetries, we have
imposed time reversal (T) invariance by requiring that a T
transformation followed by ${\bf G}^T_i$ commute (or anticommute) with
the spatial transformations. Since T sends $U_{ij} \to -U_{ij}$, we
found ${\bf G}^{T}_i = i\tau_2$ satisfies all requirements. The
combination of all these symmetries completely determines the form of
$U_{ij}$ in Eq. \ref{eq:ansatz}.

It turns out that due to enhanced symmetry in special limits, the ansatz
of Eq. \eqref{eq:ansatz} describes three different spin liquid states:
the U(1)-uniform state, the U(1)-staggered state and the Z$_2$
state. The U(1) uniform state has $\chi_{ij}>0$, no pairing
($\Delta_{ij}=0$) and a U(1) phase invariance. On the other hand, the
U(1) staggered state has no hopping $\chi_{ij}=0$ and finite
$\Delta_{ij}\neq0$ which alternates sign on adjacent triangles. In this
state, the U(1) phase invariance can be understood after noting that an
SU(2) gauge transformation can rotate this pure pairing state into a
pure hopping state (with hopping $\chi_{ij}=\Delta_{ij}$). Lastly, the
Z$_2$ state arises when both pairing and hopping are present.

In general, we need not keep the time reversal symmetry of the ansatz in
Eq. \eqref{eq:ansatz}. If we let $\chi_{ij} = u_{ij}\cos\theta_{ij}$ and
$\Delta_{ij}=u_{ij}\sin\theta_{ij}$, so that
$\text{sign}(\theta_{ij})=\text{sign}(\Delta_{ij})$, we can extend the
ansatz to
\begin{equation}
  U_{ij} = i u_{ij} \exp\{ -i \tfrac{\nu}{2}\hat n_{ij}\cdot\tau\}, \quad \vec a_i = -\mu\hat z
\end{equation}
where $\hat n_{ij} = \hat z \cos\theta_{ij} + \hat x
\sin\theta_{ij}$. This extended form then has all the same spatial
symmetries of Eq. \eqref{eq:ansatz} but recovers time reversal
invariance only at $\nu=\pi$. 

Having discussed the fermionic projective symmetry group approach to
construct a set of distinct spin liquid states on the hyper-kagome
lattice we would like to know which of them
are viable candidates for the ground state of the Heisenberg model.  
In order to address this issue, we compare the energies of these
different states.

%\bigskip

\noindent{\bf Energetics of candidate spin liquid states:} 
We have computed the ground state energy for the above class of
states in mean field theory as well as by a numerical Gutzwiller
projection of the mean field states which yields a physical spin
wavefunction. The Gutzwiller projected energy is computed using the
variational Monte Carlo (VMC) method \cite{projection}.

\begin{figure}
\includegraphics[width=0.45\textwidth,height=0.3\textwidth]{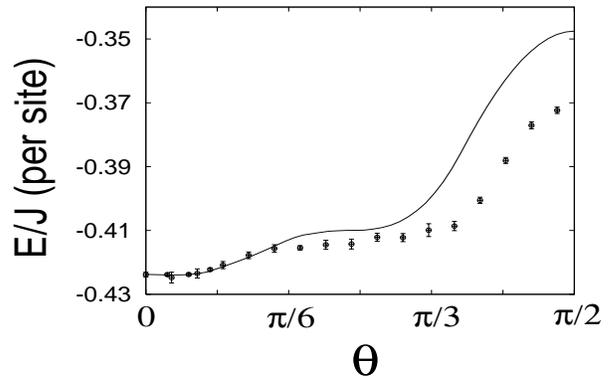}
\caption{Energetics of the Z$_2$ family of states parameterized by $\theta$.
Points show VMC data and the
solid line is the mean field result multiplied by $g_J \approx 3$.
The uniform state ($\theta=0$) appears to have the lowest energy
in this class of states.}
\label{fig:Z2energy}
\end{figure}

%\smallskip

$\bullet$ \emph{U(1)-uniform state}: The mean field ground state energy
per spin is $E^{\rm mf}_{\rm unif} = -0.144 J$. After Gutzwiller
projection, we find a variational energy $E^{\rm proj}_{\rm unif}
\approx -0.424 J$, so that $E^{\rm proj}_{\rm unif}/E^{\rm mf}_{\rm
  unif} \approx 3$. The energy of the projected state compares favorably
with the ED result on a single unit cell, $E^{\rm ed} = -0.454 J$.  In
the pre-projected state, three spinon bands cross the Fermi level. One
Fermi surface is electron-like and centered at $\vec{K} = (0,0,0)$,
while the other two are hole-like and centered about
$\vec{K}=(\pi,\pi,\pi)$. All three have $k_F \approx 0.2\pi$.
%\smallskip

$\bullet$ \emph{U(1)-staggered state}: 
The mean field energy of this state is $E^{\rm mf}_{\rm stag} = -0.122 J$. 
Due to flat bands at the chemical potential in this state, the energy
of the projected wavefunction depends somewhat on our selection of the
subset of the flat band states we fill with fermions in the preprojected
state. For various choices that we have explored the estimated VMC energy
is about $E^{\rm proj}_{\rm stag} 
\sim -0.37J$, significantly higher than the uniform state.

%\smallskip

$\bullet$ \emph{Z$_2$ state}: 
As seen from Fig. \ref{fig:Z2energy}, the mean field energy of the $Z_2$ 
states parametrized by $\theta=|\theta_{ij}|$ is higher than that of the U(1) uniform 
state (which corresponds to $\theta=0$). 
Even after projection, the U(1) uniform state appears to have
the lowest energy, although the energy is quite flat as a
function of $\theta$ for $\theta \lesssim 0.1\pi$ as in the mean field
theory. The projected energy differs from the mean
field value by about a factor of three over a wide range of $\theta$.

$\bullet$ \emph{Chiral states}: We have also checked the energetics of the
time-reversal symmetry broken chiral U(1) spin liquid ansatz. The uniform 
U(1) state is stable against such symmetry breaking. However, the 
staggered U(1) state energy is lowered by breaking time reversal 
symmetry
Nevertheless, the lowest energy obtained in this manner,
$E^{\rm proj}_{\rm chir} \sim -0.39J$, is still higher than the uniform 
U(1) state energy.

In summary, the U(1) uniform state appears to be the most favorable
candidate for the ground state of the nearest neighbor Heisenberg model
on the hyper-kagome lattice. However, as seen from Fig.~\ref{fig:Z2energy},
the energy is a rather flat function of $\theta$ for small values of
$\theta \lesssim 0.1\pi$. We therefore cannot rule out
the possibility that further neighbor couplings or an extended variational
ansatz with more variational parameters will not favor such
a $Z_2$ state with a small pair amplitude. We discuss this further in
our concluding section.

%\bigskip

\noindent{\bf Application to the specific heat of Na$_4$Ir$_3$O$_8$:}
Motivated by our variational ground state calculations we next turn to
specific heat of the uniform U(1) state in order to compare
with the data on Na$_4$Ir$_3$O$_8$. Since we cannot implement
the Gutzwiller projection exactly for computing finite temperature properties 
in any simple manner, we will use a renormalized mean field theory (RMFT) 
\cite{rmft}
to make progress. The RMFT analysis suggests that a part of 
the effect of projection
can be taken into account on top of the mean field theory by
Gutzwiller renormalization factors. For instance,
$\langle \vec{S}_i\cdot\vec{S_j}\rangle_{\rm proj}
= g_J \langle \vec{S}_i\cdot\vec{S_j}\rangle_{\rm mf}$ defines the
renormalization factor $g_J$ for the energy. This renormalizes the mean
field spinon bandwidth. From our calculations, we find 
that $E^{\rm proj}/E^{\rm mf} \approx 3$ which
implies $g_J \approx 3$.  

\begin{figure}[ht]
\includegraphics[width=0.45\textwidth]{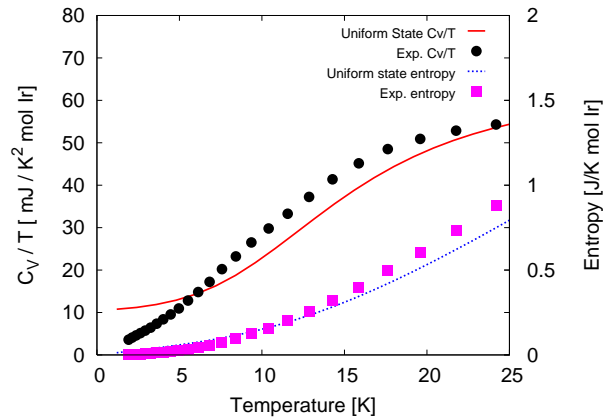}
\caption{Comparison of the specific heat coefficient, 
$C/T$, of the U(1) uniform spin liquid state computed using RMFT 
(see text) with experimental data \cite{takagi} on Na$_4$Ir$_3$O$_8$ after 
subtraction of the phonon contribution.}
\label{fig:cvplot}
\end{figure}

We next set $g_J=3$ and compute the quasiparticle contribution to the
specific heat of the U(1) uniform state in the RMFT.
Fig.~\ref{fig:cvplot} shows the heat capacity computed this way. We
emphasize that this is a zero-parameter fit. As seen, the overall
behavior of $C/T$ is in broad agreement with the data. {\it Remarkably,
  for $5K \lesssim T \lesssim 25K$, we find $C/T$ shows a strong, almost
  linear, $T$-dependence similar to experiment arising simply from the
  spinon dispersion in the uniform U(1) state}. However, as expected,
the computed $C/T$ eventually saturates below about $5K$ due to the
spinon Fermi surfaces, leading to a nonzero $\gamma \approx 10$
mJ/K$^2$/mol-Ir.  A more precise estimate of $\gamma$ requires the
projection of excited states --- this is numerically complicated due to
the many bands in this system and was not attempted in this work. We
emphasize that the broad agreement with experimental data already
provides a nontrivial check of our theory. Integrating $C/T$ we also
present a comparison between the entropy of this state which also shows
reasonable agreement with the higher temperature data.

%\bigskip

\noindent {\bf Discussion:}
We have argued, based on mean field theory and projected wavefunction
studies, that the U(1) uniform state which supports three spinon Fermi
surfaces is an energetically viable candidate for the ground state of
the $s\!=\!1/2$ hyper-kagome Heisenberg model. We have shown that it
provides a reasonable overall description of the specific heat of
Na$_4$Ir$_3$O$_8$ over a broad temperature range $T \gtrsim 5K$. Such
spin liquids with spinon Fermi surfaces have also been proposed recently
for some quasi-two-dimensional frustrated magnets \cite{spinon-fs}.  The
spinon Fermi surfaces have direct physical implications for the low
energy spin excitations.  Specifically, from the mean field Hamiltonian,
we can construct the gauge-invariant wavevectors connecting the
surfaces, which dictate the wavevector dependence of triplet $s=1$
states.

At lower temperatures, there could be Fermi surface instabilities upon
the inclusion of various small perturbations. For example, our analysis
above did not include the effect of possible further neighbor
interactions (small next neighbor antiferromagnetic exchange $J' \sim
0.1\!-\!0.2 J$ cannot be ruled out from our diagonalization study).  We
have checked that extending our $Z_2$ ansatz to include next neighbor
pairing terms, which would arise from further neighbor interactions in a
mean field theory, leads to an unconventional pairing state with line
nodes. These line nodes exist where the [110] plane (and symmetry
related planes) intersect the spinon Fermi surfaces.
Such a
line-node state would lead to a low temperature specific heat $C \sim
T^2$. As we mentioned, however, any $Z_2$ spin liquid state would most
likely undergo a phase transition to the higher temperature paramagnetic
phase, which was not observed in the experiment above $0.5K$. Moreover,
from Fig.\ref{fig:cvplot}, it seems that one need not appeal to such a
paired state to explain the linear-$T$ behavior of $C/T$ between
$5K-20K$.  If such a transition does exist, it is likely to occur at
much lower temperature. 

Finally, it has been recently argued that the weak temperature
dependence of $\chi$ at low $T$ and the specific heat $C \sim T^2$
cannot be reconciled unless spin orbit interactions are present
\cite{balents}. It was shown that, despite rather strong atomic spin
orbit coupling on Ir, the effective spin model is likely still of
Heisenberg type with the dominant effect of spin-orbit induced
Dzyaloshinskii-Moriya (DM) corrections.  For sufficiently small DM
(relative to $J$), our results for the energetics and specific heat
would remain unchanged.  However, the DM coupling can strongly effect
the spin susceptibility, especially in the line-node state. Such effects
could potentially bring the susceptibility of the line-node state, which
na\"ively behaves as $\chi(T) \sim T$, into better agreement with
experiment. The clarification of these issues is a promising direction
for future research.

We thank H. Takagi, Y. Okamoto, and P. A. Lee for helpful discussions.
This work was supported by the NSERC (A.P., M.J.L., Y.B.K.), 
CRC, CIFAR (M.J.L., Y.B.K.), an A. P. Sloan Foundation Fellowship
and an Ontario ERA (A.P.),
a David and Lucile Packard Foundation Fellowship and the NSF through 
DMR04-57440 (L.B.). 

{\it Note Added}:
During the final stages of the preparation of this manuscript, 
a paper dealing with related issues 
has appeared on arXiv \cite{zhou}.

\end{document}